\documentclass[twocolumn,superscriptaddress,pra,amsfonts,showpacs,floatfix]{revtex4}

\RequirePackage[pdftex,final]{graphicx}
\RequirePackage[utf8]{inputenc}
\RequirePackage[T1]{fontenc}

\RequirePackage[intlimits]{amsmath}
\RequirePackage{amssymb,amstext,color,xspace}
\RequirePackage{soul}

\providecommand{\abs}[1]{\left\vert #1\right\vert}
\providecommand{\lp}{\left(}
\providecommand{\rp}{\right)}
\providecommand{\lsqr}{\left[}
\providecommand{\rsqr}{\right]}
\providecommand{\lbr}{\left\{}
\providecommand{\rbr}{\right\}}

\providecommand\x{\mathbf{r}}
\providecommand\p{\mathbf{p}}
\providecommand\kv{\mathbf{k}}
\providecommand\dv{\mathbf{d}}
\providecommand\bv{\mathbf{b}}
\providecommand\qv{\mathbf{q}}
\providecommand\DV{\mathbf{D}}
\providecommand\EV{\mathbf{E}}
\providecommand\BV{\mathbf{B}}
\providecommand\PV{\mathbf{P}}
\providecommand\AV{\mathbf{A}}
\providecommand\PiV{\boldsymbol{\Pi}}
\providecommand\modefunction{\boldsymbol{\varphi}}

\newcommand{\nD}{n_\text{D}}
\newcommand{\nc}{n_\text{c}}

\newcommand{\dd}{\mathrm{d}}

\providecommand\epsNot{\varepsilon_0}

\begin{document}

\title{Depolarization shift of the superradiant phase transition}

\newcommand{\ourAddress}{\affiliation{Institute for Solid State Physics and Optics, Wigner Research Centre for Physics,\\Hungarian Academy of Sciences, P.O. Box 49, H-1525 Budapest, Hungary}}

\author{Tobias Grießer}
\ourAddress
\author{András Vukics}
\email{vukics.andras@wigner.mta.hu}
\ourAddress
\author{Peter Domokos}
\ourAddress

\begin{abstract}\noindent
We investigate the possibility of a Dicke-type superradiant phase transition of an atomic gas with an extended model which takes into account the short-range depolarizing interactions between atoms approaching each other as close as the atomic size scale, which interaction appears in a regularized electric-dipole picture of the QED of atoms. By using a mean field model, we find that a critical density does indeed exist, though the atom-atom contact interaction shifts it to a higher value than it can be obtained from the bare Dicke-model. We argue that the system, at the critical density, transitions to the condensed rather than the “superradiant” phase.
\end{abstract}


\maketitle

\section{Introduction}
Quantum electrodynamics, when applied to a large ensemble of identical atoms, leads to the surprising prediction that a so-called superradiant phase transition takes place above a critical density of the atom gas. In thermal equilibrium and also in the ground state of the system at zero temperature, the atoms in the superradiant phase develop a polarization field with non-vanishing mean value accompanied by a non-vanishing mean displacement field \cite{Hepp73,Wang73,Emary03a,Emary03b}. Originally, this prediction has been made on the ground of the Dicke model, which is admittedly oversimplified. Therefore, the idea of such a phase transition generated the immediate reaction of denying its existence and attributing it to an artifact of the incomplete modeling \cite{Rzazewski75,Emeljanov1976,Knight78,Kimura1981,Rzazewski91,Sivasubramanian2001,Bamba14}. However, the validity of the Dicke model is, in fact, surprisingly robust in treating the interaction between radiation and matter \cite{Keeling07,Vukics14}. One can define a gauge  transformation from the minimal coupling Hamiltonian – the \emph{a priori} model of atomic QED – to a regularized electric-dipole (RED) coupling Hamiltonian \cite{Vukics15} which is suitable to describe the coupling between independent atoms and well-defined modes of the radiation field, and which can then be systematically truncated to the simple form of the Dicke model.

It is essential to the Dicke model, however, that individual atoms contribute by spatially disjoint components to the polarization field; this ensures that the instantaneous atom-atom interaction can be eliminated and only the field-mediated radiative interaction is present. On the other hand, it turns out that the inter-atomic distance characteristic of the critical density obtained from the Dicke model is at the limit where the atoms can no longer be treated as independent dipoles. That is, the superradiant criticality is achieved at an atom gas density very close to the crystalline density of the given atomic species. One can even conjecture that the superradiant phase transition is nothing else but the image of condensation (liquefaction or solidification/crystallization, depending on the species) within the simplified framework of dipolar quantum electrodynamics \cite{Vukics15}. The formation of covalent bonds between atoms requires a refined description of the higher-than-dipole multipolar terms, of course, but the criticality itself can be indicated in a much simpler model  such as the dipole Hamiltonian. Nevertheless, the condition of spatially separated atoms should be released in order to investigate the radiative properties of the ensemble at the high densities considered. In this regime, the instantaneous – depolarizing – atom-atom interaction, whose range in the RED gauge is reduced to the atomic size scale \cite{Vukics15}, plays a substantial role, and is expected to shift the critical point. The physical reason of the expected shift is obvious: assume that the atoms are spontaneously polarized along a given direction in the superradiant phase. If they are allowed to approach each other on the length scale of the atomic size – something they have to do close to and above the critical coupling strength – the interaction of two dipoles pointing along the same direction costs energy and disfavors the ordered configuration.  The study of this shift is the subject of the present paper. We complement the Dicke-model description of the dense atomic gas by incorporating the contact terms – the leftover of the instantaneous atom-atom interaction in the RED gauge – accounting for the case of overlapping atoms and depolarization.

The paper is organized as follows. In Section II, we recall the necessary expressions from the RED gauge and the origin and characteristics of the depolarizing interaction in this gauge. In Section III, we make approximations beyond those leading to the Dicke model from the RED gauge Hamiltonian. These lead to a simplified model specially tailored to the study of the \emph{onset} of the superradiant phase transition. In Section IV, we look for the phase transition as a dynamical instability within this simplified model, and find the shifted critical point. We discuss our results in simple physical terms in Section V, comparing the superradiant phase transition with the commonly observed phase transition of condensation. In the Appendix, we summarize the RED gauge for the sake of completeness.

\section{Regularized electric dipole gauge}

Let us summarize the basic expressions in the gauge which has been shown to be particularly suitable to describe the quantum electrodynamics of an optically dense cloud of $N$ well-localized atoms. The RED gauge can be obtained from the standard minimal coupling gauge by canonical transformation \cite{Vukics15} as shown in the Appendix. The Hamiltonian then reads
\begin{subequations}
\begin{multline}
\label{eq:FundamentalHamiltonian}
H=\sum_{j=1}^N \lp\frac{\p_j^2}{2m}+H_{\text{e},j}\rp+U\\-\frac1\epsNot\sum_{j=1}^N \dv_j\!\cdot\!\DV(\x_j)+H_{\text{EM}}(\DV,\BV)\,,
\end{multline}
where $\x_j$ denotes the position of the atomic center of mass, $\p_j$ the corresponding momentum, and $\dv_j$ the atomic dipole moment. $H_{\text{e},j}$ denotes the internal (electronic) Hamiltonian of atom $j$, which reflects the familiar Schrödinger atom slightly perturbed by QED effects. The last term is the free electromagnetic field energy expressed in terms of the displacement instead of the electric field,
\begin{equation}
H_{\text{EM}}=\frac\epsNot2\int \lp\frac1{\epsNot^2}
\DV^2+c^2\BV^2 \rp\dd V.
\end{equation}
In the lack of free charges – that is, charges not described by the set of dipoles $\dv_j$ –, the displacement field is purely transverse: $\DV=\DV^\perp$.

The coupling between field and the atoms is linear in the atomic dipole moment and the displacement field. One of the main merits of this gauge is that the interaction between  atoms is vastly dominated by the indirect interaction via the radiation field. The key point in the present paper is, however, that we take into account also the residual instantaneous inter-atomic potential $U$. In the chosen gauge, it is composed of two terms,
\begin{equation}
U=\sum_{i\neq j}\lp U^\parallel_{i,j}+U_{i,j}^\perp\rp.
\end{equation}
\end{subequations}
The first term, $U^\parallel_{i,j}$, is just the Coulomb interaction between the charges belonging to different atoms, while 
the second term, $U_{i,j}^\perp$, cancels the strongest, dipole-dipole order of this instantaneous interaction outside a small region around the atoms with radius $\ell$ (termed the atomic “intimacy zone”). Altogether the potential $U$ is thus much weaker than the bare Coulomb interaction for separated atoms and is significant only to describe the contact interaction when an atom penetrates an other’s intimacy zone. The transverse part $U^\perp_{i,j}$ reads
\begin{subequations}
\begin{equation}
U_{i,j}^\perp(\mathbf r_{ij})=\frac1{2\epsNot}\dv_i\,K(\mathbf r_{ij})\,\dv_j,
\end{equation}
where $\mathbf r_{ij}=\mathbf r_i-\mathbf r_j$ and the matrix $K$ is given by (cf. Eq.~(\ref{eq:Kmatrix}))
\begin{equation}
K(\mathbf r)=\Gamma(r)\,\mathbb{I}-\nabla\circ\nabla\int \Gamma(r')\,G(\mathbf r-\mathbf r')\,\dd V',
\end{equation}
\end{subequations}
with $G$ the electrostatic (Dirichlet) Green’s function, $\Delta G(\mathbf r)=\delta(\mathbf r)$. The radially symmetric $\Gamma(r)$ is a regularizing (cutoff) function, it is normalized to unity, has a supporting volume of $\sim\ell^3$, with $\ell$ defined above. The precise form of $\Gamma(r)$ is immaterial \footnote{Concerning $K$ an interesting note presents itself, that sheds more light on the nature of difference between the Coulomb and the RED gauges. The origin of $U^\perp$ is certainly what we would identify as radiation in the Coulomb gauge: indeed, this is the (dipole-order) correction which cancels the instantaneous interaction to yield a fully retarded one. In the RED gauge, however, this has nothing to do with radiation since it stems from $\PV^\perp$, which is a prescribed field, cf. Eq. (\ref{eq:PtransverseK}).}.

Let us recall again the assumptions which allow the minimal coupling Hamiltonian to be transformed to the above form (for details see Ref. \cite{Vukics15} and the Appendix). One condition is that we consider only states where the internal energy of any given atom is low. Secondly, we apply an ultraviolet cutoff on the electromagnetic spectrum, that is, we discard parts with wavelengths smaller than some $\lambda_{\text{min}}\gg a_0$, where $a_0$ determines the size of an atom in the ground state (this would be the Bohr radius for hydrogen). And lastly, the length scale characterizing $\Gamma(r)$ obeys
\begin{equation}
\label{eq:Co}
a_0\ll\ell\ll\lambda_{\text{min}}.
\end{equation}

If we choose the cutoff wavelength to be of the order of the optical wavelength corresponding to the given atomic species, then there exists a range of values for $\ell$, which satisfy the above chain of inequalities \footnote{As an illustration thereof, we can quote the well-known relation for hydrogen: \[\lambda_\text{A}=\frac{16\pi}{3\alpha}\,a_0\simeq2000\,a_0.\]}.

\section{The electromagnetic and electronic subsystem}

\subsection{Adiabatic elimination of center-of-mass motion}

We shall simplify the model further by dropping the center-of-mass kinetic energy terms and regarding the atomic positions as time-independent (classical) \textit{random variables} instead. This may be viewed as a Born–Oppenheimer-type approximation, which can be justified by considering the vastly different time scales operative for the electromagnetic and electronic subsystems on the one hand, and the center-of-mass motion on the other.
Furthermore, it is not our ambition to follow the subsystem's dynamics through all times. Instead, our aim is to find the conditions under which the normal ground state of the remaining subsystem first exhibits a \textit{dynamically unstable} behavior, which point we will interpret as signaling a phase transition. Our approach thus differs from previous ones \cite{Hepp73,Wang73} in being based explicitly on dynamical as opposed to thermodynamic considerations. In order to extract the necessary information, at some point we too will need to have recourse to statistical averaging over the external degrees of freedom.

\subsection{Linearization of atomic excitation}

Since we are interested in the stability of the normal ground state of the system, we can confine the description to the lowest-lying excitations. Accordingly, we can approximate $H_{\text{e},j}$ by that of an isotropic harmonic oscillator with transition frequency $\omega$
\begin{subequations}
\label{eq:bosonization}
\begin{equation}
H_{\text{e},j}=\hbar\omega\,\bv_j^\dagger\!\cdot\!\bv_j.
\end{equation}
Accordingly, the dipole moment of the $j$th atom we can write as
\begin{equation}
\dv_j= d\lp\bv_j+\bv_j^\dagger\rp,
\end{equation}
\end{subequations}
with $d>0$ being the transition dipole. This may be viewed as an effective linearization of the theory close to the atomic ground state.

The Hamiltonian we will henceforth consider is then given by
\begin{multline}
H=\sum_{j=1}^N \hbar \omega\,\bv_j^\dagger \!\cdot\! \bv_j+\frac{d^2}{2\epsNot}\sum_{i\neq j}^N \qv_i\,u(\mathbf r_{ij})\,\qv_j+\\
-\frac{d}{2\epsNot}\sum_\nu A_\nu\sum_{j=1}^N \qv_j\!\cdot\! \modefunction_\nu(\x_j)+\sum_\nu \hbar \Omega_\nu\,a_\nu^\dagger a_\nu, \label{eq:HR}
\end{multline}
and it can be expected to describe the electronic- and electromagnetic subsystem for such a duration as is long enough for the purpose of revealing instability. Here we use the quadratures
\begin{subequations}
\begin{equation}
 \qv_j=\bv_{j}+\bv_{j}^\dagger\,,
\end{equation}
and
\begin{equation}
A_\nu=-i(a_\nu-a_\nu^\dagger).
\end{equation}
\end{subequations}
The direct atom-atom coupling matrix $u$ is given by
\begin{equation}
\label{eq:umatrix}
u(\mathbf r)=K(\mathbf r)+\nabla\circ\nabla\;G(\mathbf r).
\end{equation}
The second term represents the dipole order in the Coulomb interaction energy $U^\parallel_{i,j}$, the higher multipolar orders being neglected in accordance with the linearization in terms of the electronic degrees of freedom $\bv_j$ in Eq.~(\ref{eq:bosonization}). The supporting volume of the matrix $u$ is $\ell^3$.
Finally, we mention that we have introduced the set of transverse modes of the electromagnetic field $\modefunction_\nu$ normalized as
\begin{equation}
\label{eq:normalization}
\int\modefunction_\nu(\x)\cdot \modefunction_{\nu'}(\x)\,\dd V= 2\hbar \Omega_\nu\,\epsNot\,\delta_{\nu,\nu'}.
\end{equation}

The Hamiltonian (\ref{eq:HR}) differs from the usual single-mode Dicke Hamiltonian in several respects. Firstly, all field modes below an ultraviolet cutoff are retained; secondly, the atomic position dependence of the coupling between the atomic dipoles and the displacement field is taken into account; thirdly, the electronic degrees of freedom are represented by isotropic harmonic oscillators; lastly and most importantly, the instantaneous contact interaction energy between the atoms is accounted for (cf.~second term). This term leads to depolarization since it punishes the configuration of dipoles pointing along the same direction.

In keeping with the emphasis on dynamics, we now investigate the solutions to the equations of motion corresponding to the Hamiltonian (\ref{eq:HR}), which read
\begin{subequations}
\begin{align}\label{eq:Heisenbergb}\ddot{\qv}_{j}=&-\omega^2\, \qv_{j}-\frac{2\,d^2\omega}{\hbar\epsNot}\sum_{i\neq j}^N u(\x_{ji})\,\qv_{i}
-\frac{d\,\omega}{\hbar\epsNot}\sum_{\nu}A_\nu\modefunction_\nu(\x_j)
\\\label{eq:Heisenberga}\ddot{A}_{\nu}=&-\Omega_\nu^2 A_\nu-\frac{d\,\Omega_\nu}{\hbar\epsNot}\sum_{j=1}^N \qv_{j}\!\cdot\! \modefunction_\nu(\x_j).
\end{align}
\end{subequations}

\subsection{Coarse graining approximation}

We will now make use of the conditions (\ref{eq:Co}) and imagine the total volume $V$ divided into disjoint cells $\delta V(\x)$ centered around the points labeled $\x$, each cell being much larger than the support $\ell$ of the regularization function $\Gamma(r)$, but much smaller than the cube of the minimal wavelength, that is
\begin{equation}
\label{eq:coarseGraining}
\ell^3 \ll \delta V \ll \lambda_{\text{min}}^3.
\end{equation}

Using the first inequality – which makes that the \emph{bulk} of any two cells, even if they are neighboring, do not interact, the support of the interaction being $\ell^3$ – we neglect the instantaneous interaction of dipoles which belong to \emph{different} cells. Thus in (\ref{eq:Heisenbergb}), we restrict the second term to interaction between atoms only in the same cell, that is
\begin{equation}
\label{eq:qeq}
\ddot{\qv}_j=-\omega^2 \qv_j-\frac{2d^2\,\omega}{\hbar\epsNot}\sum_{\underset{k\neq j}{\x_k \in \delta V(\x)}} u(\x_{jk})\,\qv_k-\frac{d\,\omega}{\hbar\epsNot}\sum_{\nu}A_\nu\modefunction_\nu(\x).
\end{equation}
In the last term, we used a long wavelength approximation based on the second inequality in (\ref{eq:coarseGraining}), which makes that the mode function varies slowly on the lengthscale of a single cell.

In a similar spirit, in equation (\ref{eq:Heisenberga}) we may approximate
\begin{equation}
\sum_{j=1}^N \qv_{j}\!\cdot\! \modefunction_\nu(\x_j)\simeq  \sum_{\x}n(\x)\,\qv(\x)\!\cdot\!\modefunction_\nu(\x)\,\delta V(\x),\label{eq:Ap2}
\end{equation}
where we have introduced the cell-averaged generalized coordinate
\begin{subequations}
\begin{equation}
\qv(\x)=\frac{1}{\delta N(\x)}\sum_{\x_j\in \delta V(\x)} \qv_{j},
\end{equation}
as well as the cell density
\begin{equation}
n(\x)=\frac{\delta N(\x)}{\delta V(\x)},
\end{equation}
\end{subequations}
with $\delta N(\x)$ being the number of particles in the given cell. Later we will assume $\delta N(\x) \gg 1$ so that its statistical fluctuations be negligible.

\section{Dynamical instability in mean-field approximation}

It is at this point that we make use of statistical considerations and resort to a mean-field type approximation.
We shall restrict our attention to such dynamical modes where for all $j$ with $\mathbf r_j\in\delta V(\mathbf r)$ we can assume $\mathbf q_j \simeq \mathbf q (\mathbf r)$. The existence of such states requires that to a good approximation the sum
\begin{equation}
\sum_{\underset{k\neq j}{\x_k \in \delta V(\x)}} u(\mathbf r_{jk})\label{sum}
\end{equation}
be independent of the index $j$, i.e. the spatial configuration of dipoles surrounding any given one can be assumed to be identical on the scale $\ell$ set by the interaction, which is guaranteed if every dipole is interacting simultaneously with a large number of others within the same cell. Adopting this hypothesis, we may substitute for the sum in (\ref{eq:qeq}) its conditional expectation value
\begin{equation}
\sum_{\underset{k\neq j}{\x_k \in \delta V(\x)}} u(\mathbf r_{jk})\,\qv_k\rightarrow n\,\qv\int g(r_{jk})u(\mathbf r_{jk})\,\dd^3 x_k,
\end{equation}
wherein $g(r)$ denotes the radial distribution function of the atomic centers, which we regard as a given.
As a result we obtain
\begin{subequations}\label{eq:N1}
\begin{align}
\ddot{\qv}&=-\omega^2\lp1+\varsigma\rp\qv-\frac{d\,\omega}{\hbar\epsNot}\sum_{\nu}A_\nu\modefunction_\nu(\x),\\
\ddot{A}_\nu&=-\Omega_\nu^2A_\nu-\frac{n\,d\, \Omega_\nu}{\hbar\epsNot}\sum_\x\qv(\x)\!\cdot\!\modefunction_\nu(\x),
\end{align}
here the summation in the last term is over the cells, and we have neglected the statistical fluctuations of $\delta N$.
\end{subequations}
The contact interaction in the mean-field approximation amounts to a density-dependent transition frequency shift $\varsigma$, which is given by
\begin{equation}
\varsigma=\frac13\frac{n}{\nD}\int g(r)\;\mathrm{Tr}\lbr u(\mathbf r)\rbr\dd V.
\end{equation}
Here we have introduced the \emph{Dicke critical density}
\begin{equation}
\label{eq:DickeCritical}
\nD=\frac{\hbar \omega \epsNot}{2 d^2},
\end{equation}
which definition exactly coincides with the expression for the critical density pertaining to the original Dicke model, whence the name.

Owing to its linearity, the system (\ref{eq:N1}) may be conveniently transformed into algebraic equations by means of a Laplace transformation with the result that
\begin{equation}
A_{\nu}(s)=\frac{J_\nu(s)}{D_{\nu}(s)},
\end{equation}
where $s$ denotes the complex frequency,
\begin{equation}
D_\nu(s)=1-\frac{n}{\nD}\frac{\Omega_\nu^2\omega^2}{[s^2+\Omega_\nu^2][s^2+\omega^2(1+\varsigma)]}
\end{equation}
and $J_\nu$ subsumes all terms containing the initial conditions. The eigenfrequencies $\left\{s_\nu\right\}$ of the system are determined by the condition $D_\nu(s_\nu)=0$. One can check that, without the frequency shift $\varsigma$, the eigenfrequencies vanish just for the density $n=\nD$.

\begin{figure}
\includegraphics[width=.9\linewidth]{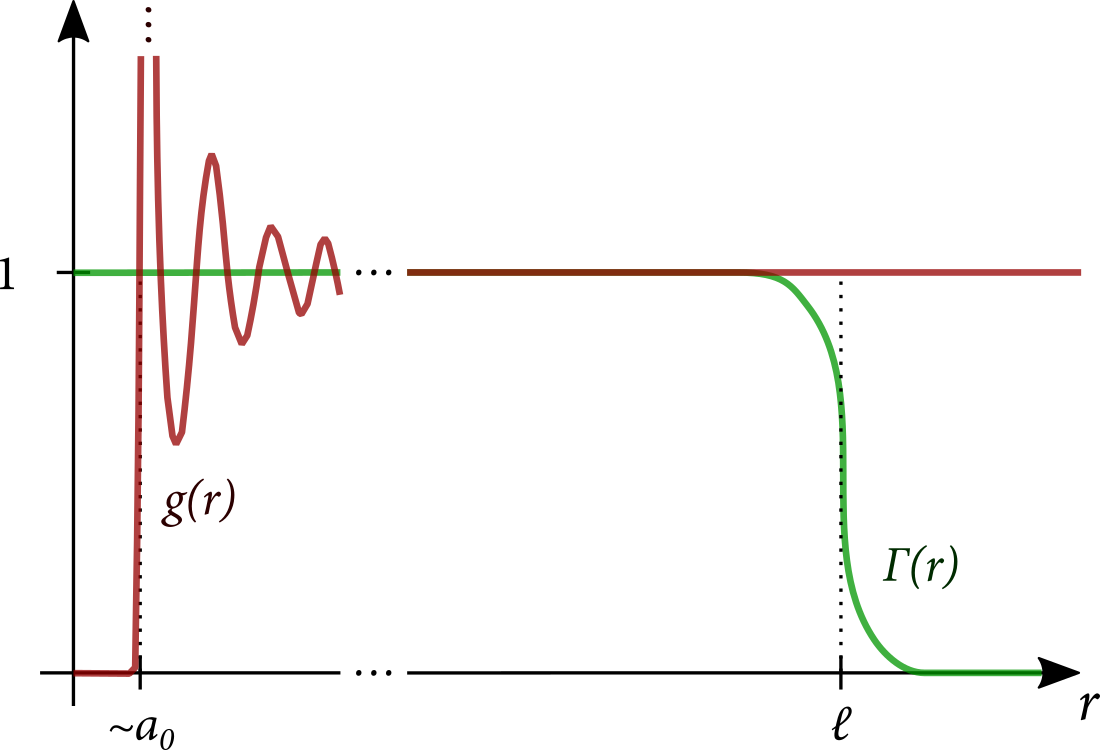}%
\caption{Relationship between the characteristic length scales of the radial distribution function $g$ (red line) and the regularizing function $\Gamma$ (green line). The former, overshooting the value 1 and oscillating on the short scale $\sim a_0$, is a generic form characteristic of hard-core repulsive interparticle potentials, with a core size $\sim a_0$. (In the case of a Lennard-Jones potential, the first peak may reach as high as 3.)}
\label{fig:sketch}
\end{figure}

To proceed, we  calculate the frequency shift $\varsigma$ from Eq.~(\ref{eq:umatrix}),
\begin{multline}
\text{Tr}\lbr u(\mathbf r)\rbr=3\Gamma(r)-\int \Gamma(r')\,\Delta G(\x-\x')\,\dd V'+\Delta G\\=2\Gamma(r)+\delta(\mathbf r),
\end{multline}
where, in the second line we have used the defining property of the Green’s function: $\Delta G(\mathbf r)=\delta(\mathbf r)$. We hence obtain
\begin{equation}
\varsigma=\frac{n}{\nD}\,\left(\frac{2}{3}\int g(r)\Gamma(r)\dd^3 x+\frac{1}{3}g(0)\right).
\end{equation}
The second term in the parentheses stems from the dipole part of the Coulomb interaction and does not in fact contribute because due to short range repulsion, we certainly have $g(0)=0$. To deal with the first term, we note that
on the scale given by $\ell$, the ensemble of atoms may be regarded as spatially uniform, because
the scale on which the radial distribution function varies around unity is clearly of the order of $a_0$. The relationship between characteristic spatial scales of $g$ and $\Gamma$ is sketched in Figure~\ref{fig:sketch}. Thus, due to (\ref{eq:Co}), we can use the normalization of $\Gamma$ to conclude that
\begin{equation}
\int g(r)\Gamma(r)\dd^3 x=1+O(a_0/\ell).
\end{equation}

Hence, to within the accuracy of the model,
\begin{equation}
\varsigma=\frac{2}{3}\frac{n}{\nD}.
\end{equation}
It is important to notice that the precise form of $\Gamma(r)$, which, as we have mentioned above, is immaterial with regard to the quality of the approximation of the model, does not influence the eigenfrequencies of the system and thus has no effect on the question of stability either. This is as it should be because we associate dynamical instability with the occurrence of an observable phenomenon, namely, a phase transition, and therefore no dependence on arbitrary quantities can be tolerated.

The eigenfrequencies of the system are found to be
\begin{equation}
 s^2_{\nu,\pm}=-S(\omega,\Omega_\nu)\pm\sqrt{S(\omega,\Omega_\nu)^2+\lp\frac13\frac n\nD-1\rp},
\end{equation}
with $S(\omega,\Omega_\nu)=\left[\omega^2(1+\varsigma)+\Omega^2_\nu\right]/(2\omega\Omega_\nu)$. Hence we find for every mode index $\nu$ that as long as $n<\nc$, where
\begin{equation}
\nc=3\,\nD,
\end{equation}
there exist two imaginary eigenfrequencies, an upper and a lower branch and the system is \emph{stable}. As the density approaches $\nc$, the lower branch $s_{\nu,+}$ softens and the system finally becomes \emph{unstable} if $n>\nc$.

Therefore, we conclude that the model predicts a phase transition at a critical density which is three times the Dicke critical density. The considerable shift of the critical density stems from the depolarization effect due to the contact interaction in the dense atomic gas.

\section{Discussion}
Let us emphasize that in the course of our analysis of the criticality in the RED Hamiltonian, we have retained all the modes of the electromagnetic field. In the present gauge, the Dicke critical density (\ref{eq:DickeCritical}) depends only on atomic parameters \footnote{The reason for this is that in the present gauge, the atom-mode coupling coefficient $g$ is proportional to the square root of the mode frequency, and hence the mode frequency drops out from the expression of the critical coupling. This is not the case in Coulomb gauge, where the coupling coefficient is proportional to the square root of the \emph{inverse} of the mode frequency.}. This means that all the modes become critical at the same point, that is, reaching the critical density of atoms, a superradiant field starts to build up in all the electromagnetic modes. In this sense, the superradiant phase transition is a macroscopic effect in the RED Hamiltonian. This is another argument for the identification of the superradiant phase transition with condensation:  the former is the silhouette of the latter in the RED framework which is too simple to capture condensation in the fullness of its radiative, electrostatic, and quantum effects. Let us recall that previously a scaling argument has been presented to support such an identification: both the covalent-bond distance and the Dicke critical distance in the RED framework depend solely on the atomic size, so there is no free parameter to tune the two separately (cf. note 32 in Ref. \cite{Vukics15}). Indeed, in the expression (\ref{eq:DickeCritical}), both $\omega$ and $d$ can be expressed by the atomic size (the Bohr radius $a_0$ for a hydrogen) to obtain
\begin{equation}
\nD=\frac1{64\,\pi\,a_0^3}.
\end{equation}

The critical density obtained with depolarization included is remarkably close to the actual condensed density. Indeed, already $\nD$ can be within the same order of magnitude as the condensed density. For example, in the case of rubidium ($D_1$ line  \cite{rsc,steck}) $\nD=1.75\times10^{27}/\text{m}^3$ \footnote{This quantity has been presented with a factor of 4 mistake in our previous paper \cite{Vukics15}.}, while the crystalline density is $11\times10^{27}/\text{m}^3$. The critical density with the depolarization interaction included is now found to be $\nc=3\,\nD=5.25\times10^{27}/\text{m}^3$. The  mean  interatomic distance corresponding to the critical density is  $17\,\text{\AA}$, whereas the atomic diameter is about
 $5\,\text{\AA}$. Although the atomic separation is still below the covalent bond distance  $4.4\,\text{\AA}$, one can expect that the higher order multipolar terms start to become significant. This can be the reason for the higher density needed for  the actual process of condensation in which the static and radiative multipoles and exchange interactions are somewhat more prohibitive to ordering than the short range depolarizing interaction in the RED framework. Of course, in the former case the ordering itself is also much more complex than the simple ferroelectric order accompanied by the build-up of superradiant fields. Nevertheless the mean field theory within the RED gauge leads to an upper density limit for the stability of the homogeneous gas phase, which is remarkably close to the condensed matter density. 

Finally, we note that currently there is a surge of interest in ultrastrong-coupling physics and the superradiant phase transition, mainly in connection with the extensive emerging field of hybrid systems \cite{Devoret2007,Geiser12,Hagenmuller2012,Todorov14a,Todorov15,Cottet2015}. In particular, in the field of circuit QED, one of the promising candidates for reaching ultrastrong coupling, even a similar debate seems to have arisen concerning the feasibility of the superradiant phase transition \cite{Nataf10,Viehmann11,Ciuti2012Comment,Jaako2016} as in the QED of atoms. It is an interesting comparison that in atomic QED, as our approach demonstrates, a very precise microscopic modeling of the atoms and their interaction with the field seems to be necessary for judging the feasibility and the nature of the superradiant phase transition.

\appendix*

\section{Derivation of the RED Hamiltonian}
\label{sec:derivation}
Our fundamental tool in the QED of atoms is a Hamiltonian obtained from a canonical transformation from the Coulomb gauge (minimal coupling) Hamiltonian, which transformation consists in a momentum shift for both field and particles. A special case of this transformation is the Power–Zineau–Woolley one, leading to the multipolar Hamiltonian \cite{Power59}, cf. also Chapter IV.C in \cite{CDG}. The transformed Hamiltonian reads (cf. Eq. (12) in Ref. \cite{Vukics14})
\begin{multline}\label{eq:Hmult}
H=
\sum_\alpha \frac{1}{2m_\alpha}\lp\p_\alpha+\frac{\partial }{\partial\x_\alpha}\int\AV\cdot\PV\,\dd{V}-q_\alpha\AV(\x_\alpha)\rp^2\\+\frac1\epsNot\int
\PiV\cdot\PV\,\dd{V}
+\frac\epsNot2\int\lp\frac{1}{\epsNot^2}
\PiV^2+c^2\lsqr\nabla\times \AV\rsqr^2 \rp\dd{V}
\\+\frac{1}{2\epsNot}\int\PV^2\,\dd{V}.
\end{multline}
Here, $\PiV$ is the canonical field momentum and $\PV$ is called “polarization field”, which incorporates the material degrees of freedom in their interaction with the EM field. It is a source of confusion, so it is not amiss noting that here we are still dealing with microscopic electrodynamics, and all this is really just a different way of describing the same things.

In earlier works \cite{Vukics14,Vukics15} we have argued that this Hamiltonian can be better suited for the quantum electrodynamics of atoms than the original, Coulomb-gauge one, for several reasons. However, it suffers from the presence of the last, the so-called $P^2$ term, that contains a distribution squared for the most straightforward choice of the polarization field, which is Power’s choice: $\PV_{\text{Power},A}(\x)=\sum_{\alpha\in A} q_\alpha\x_{\alpha}\int_0^1 \dd{u}\,\delta(\x-u\x_\alpha)$.

A solution to this problem was presented in Ref. \cite{Vukics15}, where it was pointed out that the transverse part of the polarization field $\PV$ is not uniquely defined. That is, while the longitudinal part is given unambigously by the charges via the Coulomb interaction as
\begin{subequations}\label{eq:Pol}
 \begin{multline}
  \PV^\parallel(\kv)\lp=-\epsNot\EV^\parallel(\kv)\rp\\=\frac{i}{(2\pi)^{\frac32}}\frac\kv{k^2}\sum_A e^{-i\kv\cdot\x_A}\sum_{\alpha\in A}  q_\alpha\lp e^{-i\mathbf k\cdot\delta\x_\alpha}-1\rp,
 \end{multline}
the transverse part has a certain freedom. In contrast to Power’s choice, the regularity of the transformation was imposed by choosing
\begin{equation}
\label{eq:PtransverseK}
\mathbf P^\perp(\mathbf k)=\frac{\gamma(k)}{(2\pi)^{\frac32}}\lp\mathbb{I} -\frac{\mathbf k\circ\mathbf k}{k^2}\rp\sum_A\dv_A\,e^{-i\kv\cdot\x_A},\end{equation}
\end{subequations}
where we have introduced the regularizing function $\gamma(k)$, normalized as $\gamma(0)=1$ and vanishing with $k\to\infty$. In \cite{Vukics15}, this was chosen a Lorentzian. Here the label $A$ for clusters of charges (atoms) with dipole moment $\dv_A=\sum_{\alpha\in A}q_\alpha\x_\alpha$ was introduced, while $\alpha\in A$ labels the particles belonging to cluster $A$. The position of a cluster (center of mass or atomic nucleus) is denoted by $\x_A$, while the relative positions of the constituent particles by $\delta\x_\alpha=\x_\alpha-\x_A$.

With this choice, we can find the Hamiltonian for $N$ identical atoms. Here we restrict ourselves to hydrogen-like atoms with a core and a single electron labeled by $c$ and $e$, respectively, but the extension to the general case is straightforward. There appears an asymmetry between the treatment of $\x_{c,A}$ and $\x_{e,A}$ because the former is identified with the position of the atom, and therefore it enters the expression (\ref{eq:PtransverseK}) twice, while $\x_{e,A}$ only once.
\begin{multline}\label{eq:NH}
H=\sum_A\frac1{2m_c}\left\{\mathbf p_{c,A}+q\lsqr\tilde{\AV}(\x_{c,A})-\AV(\x_{c,A})\rsqr\right.\\+\left.(\dv_A\cdot\nabla)\tilde{\AV}(\x_{c,A})+\dv_A\times \lsqr\nabla\times\tilde{\AV}(\x_{c,A})\rsqr\right\}^2\\
+\sum_A\frac{1}{2m_e}\left\{\mathbf p_{e,A}-q\lsqr\tilde{\AV}(\x_{c,A})-\AV(\x_{e,A})\rsqr\right\}^2
\\+\frac{1}{\epsNot}\sum_A
\dv_A\cdot\tilde{\mathbf \Pi}(\x_{c,A})
\\+\frac{\epsNot}{2}\int_V \left(\frac{1}{\epsNot^2}
\mathbf \Pi^2+c^2\left[\nabla\times \mathbf A\right]^2 \right)\dd^3x
\\+\sum_A\left(U_A^\Vert+U_A^\perp\right)+\sum_{A\neq B}\left(U_{A,B}^\Vert+U_{A,B}^\perp\right),
\end{multline}
where $\tilde{\AV}\equiv \gamma\star\mathbf A$ and $\tilde{\PiV}\equiv \gamma\star\PiV$ (convolution). Of the static potentials
\begin{subequations}
\begin{equation}
U_A^\Vert=\frac1{2\epsNot}\int\lp\PV_A^\Vert\rp^2\,\dd{V}=\frac1{8\pi\epsNot}\sum_{\underset{\alpha\neq\beta}{\alpha,\beta\in A}}\frac{q_\alpha\,q_\beta}{\abs{\x_\alpha-\x_\beta}}
\end{equation}
is the intra-atomic and
\begin{equation}
U_{A,B}^\Vert=\frac1{2\epsNot}\int\PV_A^\Vert\cdot\PV_B^\Vert\,\dd{V}=\frac1{8\pi\epsNot}\sum_{\alpha\in A,\beta\in B}\frac{q_\alpha\,q_\beta}{\abs{\x_\alpha-\x_\beta}}
\end{equation}
\end{subequations}
is the inter-atomic Coulomb potential while the terms stemming from the perpendicular part of the $P^2$ term read
\begin{subequations}
\begin{equation}
\label{eq:perturbing}U_A^\perp=\frac1{2\epsNot}\int\lp\PV_A^\perp\rp^2\,\dd{V}=\frac{q^2r^2}{3\epsNot}\Gamma(0),
\end{equation}
$r$ being the position operator of the valence electron relative to the core, and
\begin{equation}
U_{A,B}^\perp=\frac1{2\epsNot}\int\PV_A^\perp\cdot\PV_B^\perp\,\dd{V}=\frac{1}{2\epsNot}\dv_A\,K(\x_A-\x_B)\,\dv_B,
\end{equation}
\end{subequations}
where $\Gamma\equiv \gamma\star \gamma$ and this function is, just like $\gamma$, normalized to unity, i.e.
\begin{equation}
\int \Gamma(r)\,\dd^3x=\int \gamma(r)\,\dd^3x=1.
\end{equation}
We have also introduced the matrix (relying on the Parseval-Plancherel identity)
\begin{multline}
\label{eq:Kmatrix}
K(\x)\equiv\int\dd^3 k\,\frac{\gamma^2(k)}{(2\pi)^3}\lp\mathbb{I}-\frac{\kv\circ\kv}{k^2}\rp e^{i\,\kv\cdot\x}\\=\int\dd^3 k\,\frac{\gamma^2(k)}{(2\pi)^3}e^{i\,\kv\cdot\x}+\nabla\circ\nabla\int\dd^3 k\,\frac{\gamma^2(k)}{(2\pi)^3}\frac{e^{i\,\kv\cdot\x}}{k^2}\\=\Gamma(r)\,\mathbb{I}-\nabla\circ\nabla\int \Gamma(r')\,G(\x-\x')\,\dd^3 x',
\end{multline}
with the electrostatic (Dirichlet) Green’s function
\begin{equation}
G(\x)\equiv-\frac{1}{4\pi\abs{\x}}.
\end{equation}
The first term in the last line of (\ref{eq:Kmatrix}) is a (regularized) contact interaction, while the physical effect of the second term is that $U_{A,B}^\perp$ knocks out the dipole order from $U_{A,B}$ outside of the intimacy region of the atoms.

The two terms in the second line of Eq.~(\ref{eq:NH}) are magnetic terms, and the difference between the canonical and kinetic momenta in the new picture is also a magnetic term (these are the so-called Röntgen terms, cf.~Section IV.C.4.c in \cite{CDG}). These terms are of the same order of magnitude as the electric quadrupole, and hence are neglected in the (regularized) \emph{electric-dipole} gauge.

Let us denote the length scale which characterizes the size of the support of $\Gamma$ (and $\gamma$) in real space by $\ell$. Notice that for those modes whose wavelengths are much smaller than $\ell$, we have
$\tilde{\mathbf \Pi}\sim0\sim\tilde{\mathbf A}$ and consequently the coupling of the atoms to such modes is the same as in the minimal coupling Hamiltonian. On the other hand, for modes with a wavelength much larger than the above scale, we have the usual dipole coupling, because $\tilde{\mathbf \Pi}\sim \mathbf \Pi$ and $\tilde{\mathbf A}\sim \mathbf A$, i.e. $\gamma$ acts like a delta function on such scales.
As a basic requisite of the theory, we would like to ensure that the intra-atomic low-energy spectrum is negligibly perturbed with respect to the Coulomb one (cf. Eq.~(\ref{eq:perturbing})). It can be shown that this requires $\ell \gg a_0$, where $a_0$ determines the size of the atom given only Coulomb interactions between core and electron (Bohr radius). As a second simplification, we disregard the electromagnetic spectrum below a certain cutoff wavelength $\lambda_{\text{min}}$. The coupling to the remaining part of the spectrum should be given by the usual dipole Hamiltonian.
Taken together, the requirements are given by
\begin{equation}
a_0\ll \ell \ll\lambda_{\text{min}}\label{eq:FundCond}
\end{equation}
and the (low-energy) Hamiltonian then approximately reads
\begin{subequations}
\begin{multline}
H=\sum_A\frac{\mathbf p_{c,A}^2}{2m_c}+\sum_A H_{e,A}
\\+\sum_{A\neq B}(U_{A,B}^\Vert+U_{A,B}^\perp)+\frac{1}{\epsNot}\sum_A
\dv_A\cdot\mathbf \Pi(\x_{c,A})
\\+\frac{\epsNot}{2}\int \left(\frac{1}{\epsNot^2}
\mathbf \Pi^2+c^2\left[\nabla\times \mathbf A\right]^2 \right)\dd^3x,
\end{multline}
with the electronic Hamiltonian
\begin{equation}
H_{e,A}=\frac{\mathbf p_{e,A}^2}{2m_e}+U_A^\Vert(\x_{e,A})+U_A^\perp(\x_{e,A}).
\end{equation}
\end{subequations}

\begin{acknowledgments}
This work was supported by the Hungarian Academy of Sciences (Lend\"ulet Program, LP2011-016) and the National Research, Development and Innovation Office (K115624). A.V. acknowledges support from the János Bolyai Research Scholarship of the Hungarian Academy of Sciences.
\end{acknowledgments}



\end{document}